\documentclass[noshowpacs,amsmath,twocolumn,
aps,prb]
{revtex4-1}

\usepackage{setspace}
\usepackage{amsmath}
\usepackage{graphicx}
\usepackage[nearskip,margin = 0pt]{subfig}
\usepackage{subfig}

\usepackage{verbatim}
\usepackage{amsfonts}
\usepackage{amssymb}
\usepackage{epstopdf} 
\usepackage{xcolor}
\DeclareGraphicsExtensions{.pdf,.eps,.png,.jpg,.mps} 

\begin{document}

\title{ Microresonator Soliton Dual-Comb Spectroscopy }

\author{Myoung-Gyun Suh$^{1,\dagger}$, Qi-Fan Yang$^{1,\dagger}$, Ki Youl Yang$^1$, Xu Yi$^1$, and Kerry J. Vahala$^{1,*}$\\
$^1$T. J. Watson Laboratory of Applied Physics, California Institute of Technology, Pasadena, California 91125, USA.\\
$^\dagger$ These authors contributed equally to this work.\\
$^*$Corresponding author: vahala@caltech.edu
}

\maketitle
\newcommand{\ts}{\textsuperscript}
\newcommand{\tsb}{\textsubscript}


{\bf \noindent Rapid characterization of optical and vibrational spectra with high resolution can identify species in cluttered environments and is important for assays and early alerts. In this regard, dual-comb spectroscopy has emerged as a powerful approach to acquire nearly instantaneous Raman and optical spectra with unprecedented resolution. Spectra are generated directly in the electrical domain and avoid bulky mechanical spectrometers. Recently, a miniature soliton-based comb has emerged that can potentially transfer the dual-comb method to a chip platform. Unlike earlier microcombs, these new devices achieve high-coherence, pulsed mode locking. They generate broad, reproducible spectral envelopes, which is essential for dual-comb spectroscopy. Here, dual-comb spectroscopy is demonstrated using these devices. This work shows the potential for integrated, high signal-to-noise spectroscopy with fast acquisition rates.}

Since their demonstration in the late 1990s \cite{jones2000carrier,holzwarth2000optical,diddams2001optical}, optical frequency combs have revolutionized precision measurements of time and frequency and enabled new technologies such as optical clocks\cite{diddams2001optical} , low-noise microwave generation \cite{fortier2011generation} and dual-comb spectroscopy\cite{schiller2002spectrometry,keilmann2004time,coddington2008coherent,coddington2010coherent,giaccari2008active,bernhardt2010cavity,ideguchi2014adaptive,okubo2015ultra}, while also adding performance capablibility to methods like coherent LIDAR \cite{minoshima2000high,ye2004absolute,swann2006frequency}. In spectroscopic applications, frequency comb systems exist across a broad spectral range spanning ultraviolet to infrared, making them well suited for measurement of diverse molecular species. At the same time, the method of dual-comb spectroscopy leverages the coherence properties of combs for rapid, broad-band spectral analysis with high accuracy\cite{coddington2016dual}. 

\begin{figure}
    \begin{centering}
        \includegraphics[width=1.25\linewidth]{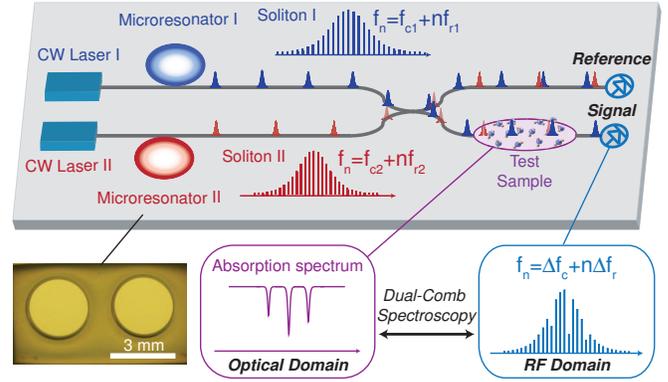}
        \captionsetup{singlelinecheck=no, justification = RaggedRight}
        \caption{{\bf Microresonator-based dual-comb spectroscopy}. Two soliton pulse trains with slightly different repetition rates are generated by continuous optical pumping of two microresonators. The pulse trains are combined in a fiber bidirectional coupler to produce a signal output path that passes through a test sample as well as a reference output path. The output of each path is detected to generate an electrical interferogram of the two soliton pulse trains. The interferogram is Fourier transformed to produce comb-like electrical spectra having spectral lines spaced by the repetition rate difference of the soliton pulse trains. The absorption features of the test sample can be extracted from this spectrum by normalizing the signal spectrum by the reference spectrum. Also shown is the image of two, silica wedge disk resonators. The disks have a 3 mm diameter and are fabricated on a silicon chip.}
    \end{centering}
\end{figure}

In parallel with advancements in frequency comb applications,  the past decade has witnessed the appearance of a miniature optical frequency comb or microcomb\cite{del2007optical,kippenberg2011microresonator}.  These microcombs have been demonstrated across a range of emission bands using several dielectric materials\cite{del2007optical,savchenkov2008tunable,grudinin2009generation,papp2011spectral,okawachi2011octave,li2012low,hausmann2014diamond}. Under continuous-wave laser pumping, the combs are initiated by way of parametric oscillation \cite{kippenberg2004kerr,savchenkov2004low} and are broadened by cascaded four-wave mixing\cite{del2007optical, kippenberg2011microresonator} to spectral widths that can encompass an octave of spectrum\cite{okawachi2011octave}. Four-wave mixing in the ultra-fast intraband gain medium of quantum cascade lasers (QCL) has also been shown to create frequency modulation (FM) combs\cite{hugi2012mid}. These FM systems have been applied to demonstrate dual-comb spectroscopy in the mid infrared\cite{villares2014dual}. Also, heterodyne of two conventional microcombs in the mid infrared has been demonstrated, a key step towards dual-comb spectroscopy\cite{yu2016silicon}. Direct heterodyne detection of two QCL FM combs in the laser current has also been shown\cite{rosch2016chip}. 

\begin{figure*}
    \begin{centering}
        \includegraphics[width=\linewidth]{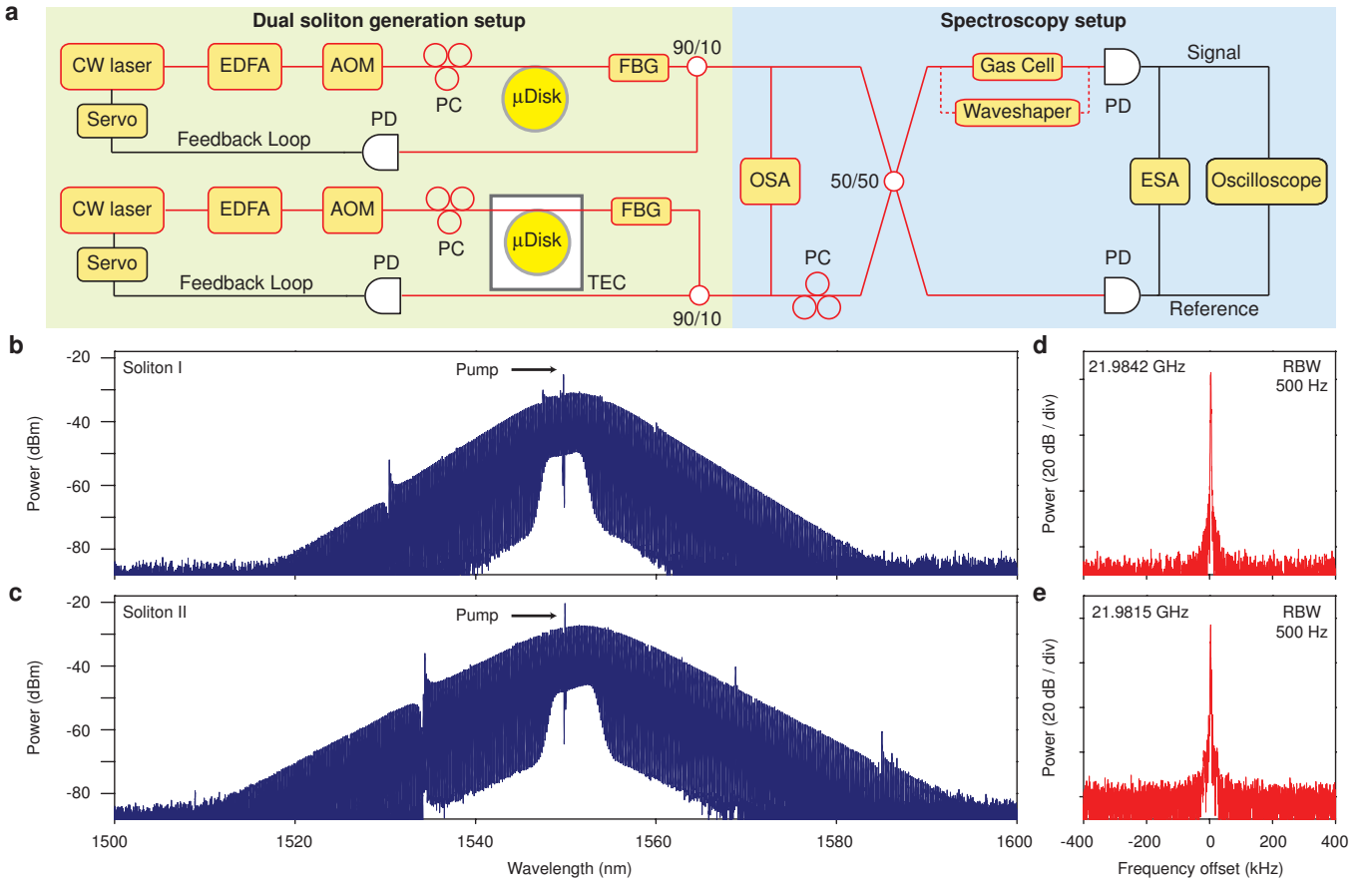}
        \captionsetup{singlelinecheck=no, justification = RaggedRight}
        \caption{{\bf Detailed experimental setup and soliton comb characterization}. {\bf a,} Continuous-wave (CW) fiber lasers are amplified by erbium-doped fiber amplifiers (EDFA) and coupled into high-Q silica wedge microresonators via tapered fiber couplers. An acousto-optic modulator (AOM) is used to control pump power to trigger soliton generation in the microresonators. Polarization controllers (PC) are used to optimize resonator coupling. A fiber Bragg grating (FBG) removes the transmitted pump power in the soliton microcomb. The pump laser frequency is servo controlled to maintain a fixed detuning from the microcavity resonance by holding the soliton average power to a fixed setpoint. An optical spectrum analyzer (OSA) monitors the spectral output from the microresonators. The two soliton pulse streams are combined in a bidirectional coupler and sent to a gas cell (or a WaveShaper) and a reference path. The interferograms of the combined soliton pulse streams are generated by photodetection (PD) and recorded on an oscilloscope. The repetition rates of the soliton pulse streams are also monitored by an electrical spectrum analyzer (ESA). The temperature of one resonator is controlled by a thermoelectric cooler (TEC) to tune the optical frequency difference of the two solitons. {\bf b-c,} Optical spectra of the microresonator soliton pulse streams.  {\bf d-e,} Electrical spectra showing the repetition rates of the soliton pulse streams. The rates are given in the legends.}
\end{centering}
\end{figure*}

\begin{figure*}
  \begin{centering}
    \includegraphics[width=\linewidth]{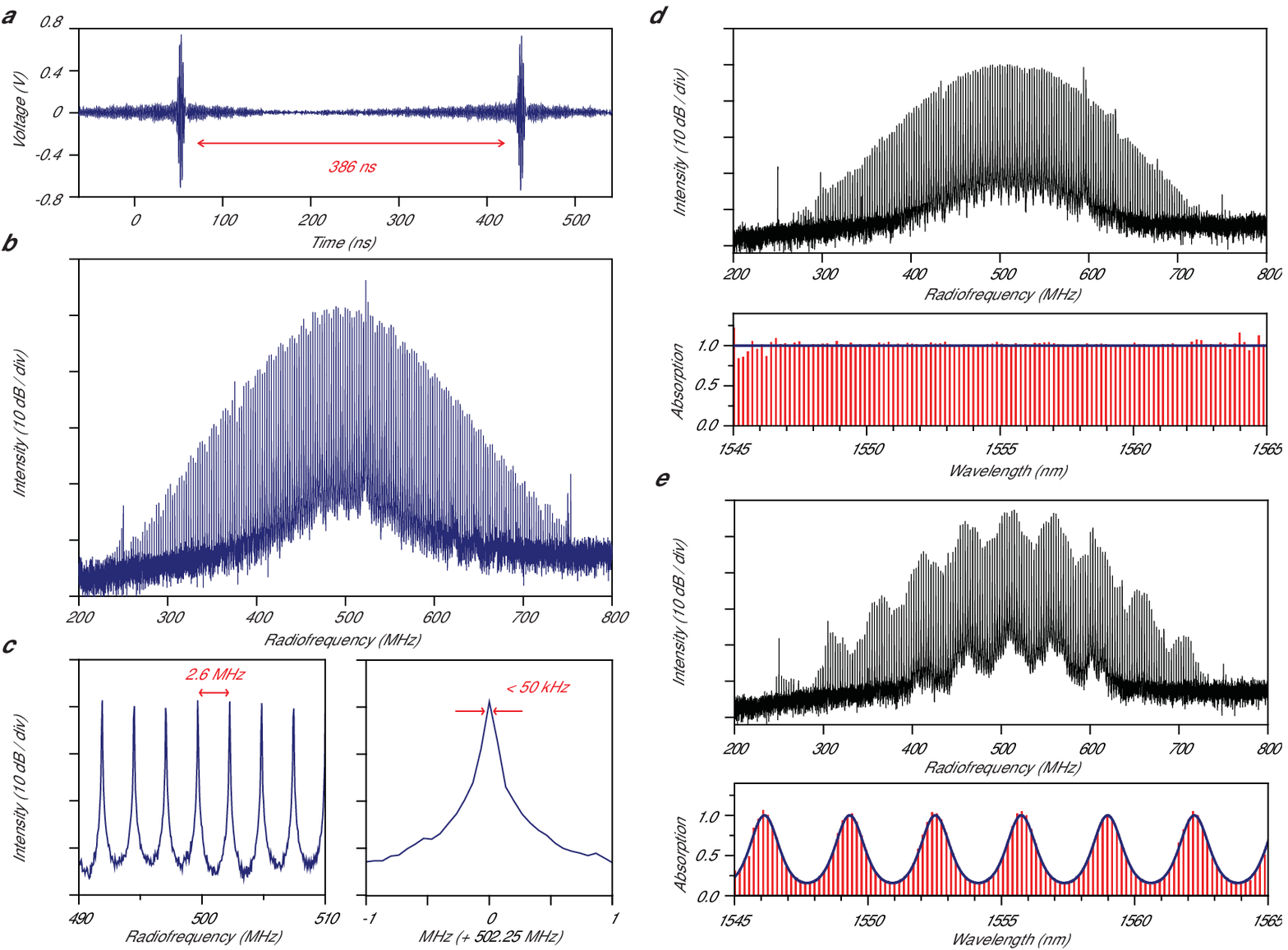}
    \captionsetup{singlelinecheck=no, justification = RaggedRight}
    \caption{{\bf Measured electrical interferogram and spectra.} {\bf a,} The detected interferogram of the reference soliton pulse train. {\bf b,} Typical electrical spectrum obtained by Fourier transform of the temporal interferogram in 3a. To obtain the displayed spectra, ten spectra each are recorded over a time of 20 $\mathrm{\mu}$s and averaged. {\bf c,} Resolved (multiple and individual) comb lines of the spectrum in 3b are equidistantly separated by 2.6 MHz, the difference in the soliton repetitation rate of the two microresonators. The linewidth of each comb line is $<$ 50 kHz and set by the mutual coherence of the pumping lasers. {\bf d-e,} Fourier-transform (black) of the signal interferogram produced by coupling the dual-soliton pulse trains through the WaveShaper (see Fig. 2a) with programmed absorption functions (spectrally flat and sine-wave). The obtained dual-comb absorption spectra (red) are compared with the programmed functions (blue curves) from 1545 nm to 1565 nm.}
    \label{fig3}
  \end{centering}
\end{figure*}

A major advancement in microcombs has been the realization of soliton mode-locking\cite{herr2014temporal,yi2015soliton,brasch2016photonic,wang2016intracavity,joshi2016thermally}. Soliton microcombs feature dissipative Kerr solitons that leverage the Kerr nonlinearity to both compensate dispersion and to overcome cavity loss by way of parametric gain\cite{herr2015dissipative}.  Unlike earlier microcombs, this new device provides phase-locked femtosecond pulses with well-defined, repeatable spectral envelopes, which is important for dual-comb spectroscopy. Their pulse repetition rate is detectable and has excellent phase noise characteristics\cite{yi2015soliton}.  In this work, we demonstrate dual-comb spectroscopy using this new platform. The dual-comb source spans over 30 nm with 22 GHz optical spectral resolution and the interferogram spectra feature high signal-to-noise. Also, precise microfabrication control enables close matching of the repetition rates so that over 4 THz of optical bandwidth is measured within 500 MHz of electrical bandwidth. 

A schematic view of the dual comb experimental setup is provided in Fig. 1. Two soliton trains having different repetition rates ($\Delta f_{r} = f_{r1} - f_{r2}$) are generated from distinct microresonators and then combined using a directional coupler. One of the combined streams is coupled through a gas cell of molecules whose absorption spectrum is to be measured. The other combined stream functions to provide a reference. The slight difference in repetition rates of the soliton streams creates a periodically time-varying interferogram in the detected current with a period $1/ \Delta f_{r}$.  Fourier transform of this time-varying signal reveals the interfering soliton comb spectra, now shifted to radio-frequency rates. The signal spectrum containing the molecular absorption information is then normalized using the reference spectrum to reveal the spectral absorption of the gas cell. 

Figure 2a gives further details on the experimental setup. Solitons are generated and stabilized in two microresonators using the active-capture/locking technique\cite{yi2016active}. The microresonators are pumped at 1549.736 nm and 1549.916 nm using two amplified fiber lasers (Orbits Lightwave), but in principle, pumping from a single laser is possible. The difference frequency of the pumps was determined to be 22.5 GHz by detecting their electrical beat note and measurement on a spectrum analyzer. After amplification, each pump laser is coupled to an acousto-optic frequency modulator (AOM). The frequency-shifted output of the AOM is used to provide a controllable optical pumping power that is required for soliton triggering\cite{yi2016active}. The pump light is then evanescently coupled into the silica microresonator via a fiber taper\cite{cai2000observation,spillane2003ideality}. Residual pumping light that is transmitted past each resonator is filtered using a fiber Bragg grating (FBG).  After the FBG, a 90/10 tap is used to monitor the soliton power for feedback control of the pump laser frequency so as to implement soliton locking\cite{yi2016active}. The optical spectra of the individual soliton streams was monitored using a Yokogawa optical spectrum analyzer. Additional precision calibration of the spectra was possible using a Wavelength References Clarity laser locked to a molecular absorption line. Typical soliton optical spectra are presented in Figs. 2b-c and feature the characteristic $\mathrm{sech}^2$ envelope observed in this case over a 60 nm wavelength span. The detected electrical spectrum for each soliton source is also shown in Figs. 2d-e. The narrow spectral lines measured with a resolution bandwidth of 500 Hz have a signal-to-noise greater than 75 dB showing that corresponding repetition rates are extremely stable. 

The high-Q resonators used in this work are described elsewhere\cite{lee2012chemically}. Briefly, they are silica wedge devices fabricated on a silicon wafer using a combination of lithography and wet/dry etching. The unloaded quality factor of the microresonators exceeds 300 million, and the generated solitons have repetition rates determined primarily by the diameter of the devices (3 mm). The repetition rate difference of the two microcomb devices is controlled by varying the silica resonator etching time\cite{lee2012chemically}.   

The optical outputs from the stabilized soliton sources are combined and coupled into two paths as shown in Fig. 2a. One path contains a 16.5 cm-long 300 Torr H\ts{13}CN gas cell manufactured by Wavelength References, Inc. which functions as the test sample in the measurement. The other path is coupled directly to a photodetector and functions as the reference. The test sample path also contained an alternate path in which a Finisar WaveShaper was inserted. The WaveShaper required an erbium fiber amplifier to compensate its insertion loss. As demonstrated below, the WaveShaper allowed synthesis of arbitray spectral transmission profiles to further verify the dual comb operation. Detection to generate the interferograms used u2t photodetectors with bandwidths of 50 GHz. Temperature control of one of the microresonators was used to tune the relative optical frequency difference of the two solitons streams. In the measurements this difference was held below 1 GHz, allowing the observation of the temporal interferogram on an oscilloscope (bandwidth 1 GHz). The spectrum of the photocurrent signals was also measured to determine the soliton repetition rates (see Figs. 2d-e) using an electrical spectrum analyzer (Rhode Schwartz) with a bandwidth of 26 GHz. 

\begin{figure}
  \begin{centering}
    \includegraphics[width=\linewidth]{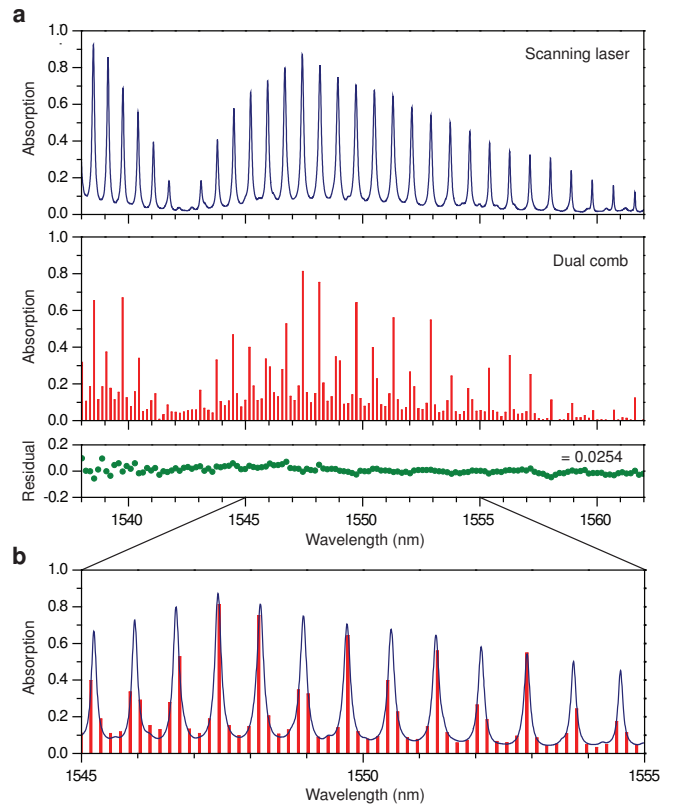}
    \captionsetup{singlelinecheck=no, justification = RaggedRight}
    \caption{{\bf Measured molecular absorption spectra. }{\bf a,} Absorption spectrum of 2$\nu_3$ band of H$^{13}$CN measured by direct power transmission using a wavelength-calibrated scanning laser (see Methods section) and comparison to the microresonator-based dual-comb spectrum. The residual difference between the two spectra is shown in green. {\bf b,} Overlay of the directly measured optical spectrum and the dual-comb spectrum showing line-by-line matching. The vertical positions of the two spectra are adjusted to compensate insertion loss.}
    \label{fig4}
  \end{centering}
\end{figure}

The reference interferogram produced by detection of the lower path in Fig. 2a and as recorded on the oscilloscope is shown in Fig. 3a. It has a period of 386 ns, corresponding to a soliton repetition rate difference of 2.6 MHz. This relatively small repetition rate difference was possible by precise lithographic control of the 22 GHz soliton repetition rate. It was possible to fabricate disks with even more closely matched repetition rates ($<$ 100kHz).  Figure 3b shows the calculated Fourier transform of the interferogram. The small repetition rate difference on the much larger 22 GHz soliton repetition rate makes it possible compress an optical span of 4 THz (1535 nm to 1567 nm) into 500 MHz of electrical spectrum. The measured wavelength span is actually narrower than the observable wavelength span of the original soliton pulse streams and is limited by the photodetector noise. The interferogram spectrum has a signal-to-noise ratio (SNR) in excess of 30 dB near the central lines. A zoom-in of the spectrum (multi- and single-line) is provided in Fig. 3c. The electrical comb lines are equidistantly separated by 2.6 MHz and have a full-width-half-maximum linewidth less than 50 kHz, limited by the mutual coherence of the independent fiber pump lasers. The pump laser line in a dissipative Kerr soliton is also a comb tooth in the soliton optical spectrum. As a result, the frequency jitter in each pump is transferred as an overall shift on the resulting soliton comb. Externally locking the two combs should reduce the observed linewidth in the interferogram spectum. 

It is interesting to note the placement of the pump lines toward the high frequency side (near 550 MHz) of the spectral maximum in the interferogram spectrum (see Fig. 3b). In the optical spectra (Figs. 2b-c) the pump is blue detuned relative to the soliton spectral maximum (this occurs on account of the Raman self-frequency-shift of the soliton\cite{Yi:16,karpov2016raman,milian2015solitons}). This spectral landmark shows that the relative spectral placement of the soliton combs is such that high optical frequencies are mapped to high interferogram frequencies. It is also interesting to note how certain non-idealities in the soliton spectra map into the interferogram spectrum. Specifically, there are avoided-mode-crossing induced Fano-like spurs\cite{yi2015soliton} in the soliton optical spectra (Figs. 2b-c) occurring near 1535 nm and this generates a corresponding feature at 750 MHz in Fig. 3b.

As an initial test of the dual-comb source, artificial absorption spectra were programmed in a Finisar WaveShaper 1000S and then measured as dual-comb spectra. In Figs. 3d-e, electrical spectra Fourier-transformed from the signal interferograms after coupling through the WaveShaper are shown. The two programmed functions are a spectrally flat 3 dB absorption and a sine-wave absorption having a 4 dB amplitude. The absorption spectra, obtained by normalizing the signal and reference electrical spectra, are compared with the programmed functions in Figs. 3d-e. The ability to reconstruct these synthetic spectral profiles clearly demonstrates the reproducibility of solitonic spectral profile. 

Finally, the absorption spectrum of the H\ts{13}CN 2$\nu_3$ band is studied. In Fig. 4a, the measured dual-comb absorption spectrum from 1538 nm to 1562 nm is shown in red and compared with a directly measured absorption spectrum shown in blue. Both absorption spectra are normalized. The direct measurement was performed by a wavelength-calibrated scanning laser (see Methods section). Sampling-induced choppiness of the dual-comb spectrum is caused by the relatively coarse spectral resolution of the solitons in comparison to the spectral scale of the H\ts{13}CN absorption lines. Nonetheless, the characteristic envelope of H\ts{13}CN 2$\nu_3$ band is clearly resolved. The residual difference between the two absorption spectra is shown in green and the calculated standard deviation is 0.0254. Furthermore, a line-by-line overlay of the measured optical and dual-comb spectra is shown in Fig. 4b to visually confirm the wavelength precision and absorption intensity accuracy of the dual-comb source.    

In principle, a finer comb spacing (lower repetition frequency) soliton source is possible. Non-soliton microcombs having mode spacings as narrow as 2.4 GHz have been demonstrated using the silica wedge resonator platform \cite{li2012low}. Modulating the microcombs by an integer factor of the repetition frequency using electro-optical modulators is another possible way to create a finer spectral comb grid. On the other hand, larger mode spacings could allow studies of fast dynamic processes such as chemical reactions and rapid measurements of the broad absorption features in liquids or solids \cite{fleisher2014mid,reber2016cavity}.

In conclusion, two soliton microcombs featuring highly balanced microwave repetition rates were used as a dual-comb spectroscopy system to measure the absorption spectrum of the 2$\nu_3$ band of H$^{13}$CN. This is the first demonstration of a microresonator soliton-based dual-comb spectroscopy system. The dual-comb source has a high SNR and spans over 30 nm in optical C-band. Using fiber nonlinear broadening or internal (resonator) dispersive wave generation, it should be possible to greatly extend this spectral span \cite{jost2015counting,okubo2015ultra}. With careful engineering of the resonator dispersion \cite{yang2016broadband} it should also be possible to cover other spectral ranges within the transmission window of silica. More generally, a wide range of materials are available for microcombs enabling access to mid infrared spectra.  With further improvements, it should also be possible to realize chip-based dual-comb coherent anti-Stokes Raman spectroscopy (CARS). The integration with other on-chip devices\cite{yang2016ultra} makes soliton-based microcombs well suited for possible realization of a dual-comb spectroscopic system-on-a-chip.

\vspace{3 mm}

\noindent\textbf{Methods}

\noindent The H$^{13}$CN absorption spectrum in Fig. 4a is obtained by coupling an external cavity diode laser (ECDL) into the H$^{13}$CN gas cell and scanning the laser while monitoring the transmitted optical power. A separate signal is also tapped from the ECDL to function as a reference. The relative wavelength change of the ECDL during the scan is calibrated using a fiber Mach-Zehnder interferometer and absolute calibration is obtained using a reference laser which is locked to a molecular absorption line (Wavelength References Clarity laser). The signal passing through the gas cell and the reference transmissions are recorded simultaneously, and the absorption spectrum in Fig. 4a is extracted by dividing the signal transmission by the reference transmission.

\vspace{3 mm}

\noindent \textbf{Acknowledgments} The authors thank Nathan Newbury at the National Institute of Standards and Technology (NIST) and Giacomo Scalari at ETH Zurich for helpful comments on this manuscript. The authors gratefully acknowledge support from the Defense Advanced Research Projects Agency (DARPA) under the PULSE and SCOUT programs, the National Aeronautics and Space Administration (NASA) and the Kavli Nanoscience Institute (KNI).

\vspace{1 mm}

\noindent\textbf{Author Contributions} Experiments were conceived by all co-authors. Analysis of results was conducted by MGS, QFY and KJV. MGS and QFY performed measurements with assistance from XY. KYY fabricated devices. All authors participated in writing the manuscript. 
\vspace{1 mm}

\noindent \textbf{Author Information} Correspondence and requests for materials should be addressed to KJV (vahala@caltech.edu ).

\bibliography{main}

\end{document}